\documentclass[11pt]{article}
\usepackage{amsmath, amssymb}
\begin{document}
\renewcommand{\today}{}
\newcommand{\gff}{generalized free field}
\newcommand{\set}{stress-energy tensor}
\newcommand{\klw}{K\"allen-Leh\-mann weight}
\newcommand{\poi}{Poincar\'e}
\newcommand{\ADS}{{\mathbb{A}{\scriptstyle\mathbb{D}}\mathbb{S}}} 
\newcommand{\PADS}{{\rm P}\ADS} 
\newcommand{\ads}{{\rm AdS}} \newcommand{\kg}{{\rm KG}} 
\newcommand{\tot}{{\rm tot}} \newcommand{\dual}{{\rm dual}}
\newcommand{\cft}{{\rm CFT}} \newcommand{\inv}{^{-1}}
\newcommand{\Eins}{{\mathbf 1}}
\newcommand{\NN}{{\mathbb{N}}} \newcommand{\MM}{{\mathbb{M}}} 
\newcommand{\RR}{{\mathbb{R}}} \newcommand{\CM}{{\mathbb{CM}}}
\newcommand{\HH}{{\mathcal H}}
\newcommand{\be}{\begin{equation}} \newcommand{\ee}{\end{equation}}
\newcommand{\bea}{\begin{eqnarray}}\newcommand{\eea}{\end{eqnarray}}
\newcommand{\QED}{\hspace*{\fill}Q.E.D.\vskip2mm}
\renewcommand{\theequation}{\thesection.\arabic{equation}}
\title{\bf Generalized free fields \\ and the AdS-CFT correspondence}  
\author{{\sc Michael D\"utsch}\thanks{Electronic address: 
  \tt duetsch@theorie.physik.uni-goe.de} {} and 
{\sc Karl-Henning Rehren}\thanks{Electronic address:
{\tt rehren@theorie.physik.uni-goe.de}} \\[2mm]
Institut f\"ur Theoretische Physik, Universit\"at G\"ottingen,
\\ 37073 G\"ottingen, Germany }

\maketitle

\begin{abstract} Motivated by structural issues in the AdS-CFT
  correspondence, the theory of generalized free fields is
  reconsidered. A stress-energy tensor for the \gff\ is constructed as
  a limit of Wightman fields. Although this limit is singular, it
  fulfils the requirements of a conserved local density for the
  \poi\ generators. An explicit ``holographic'' formula relating
  the Klein-Gordon field on AdS to \gff s on Minkowski space-time is
  provided, and contrasted with the ``algebraic'' notion of holography. 
  A simple relation between the singular \set\ and the canonical AdS
  \set\ is exhibited. 
\end{abstract}

PACS 2001: 11.10.-z, 04.62.+v


\section{Introduction}
\setcounter{equation}{0}
According to Maldacena's conjecture \cite{M}, type IIB string theory on 
5-dimensional asymptotically Anti-deSitter (AdS) backgrounds with five
compactified dimensions is 
equivalent to a maximally supersymmetric Yang-Mills theory in
physical Minkowski space-time. In the limit of large number of colors 
$N \to \infty$ and large 't~Hooft coupling $\theta = Ng^2 \to \infty$,
it is conjectured that the string may be replaced by classical 
supergravity on AdS. Roughly speaking, $1/N$ corrections correspond to
quantum corrections, and the strong coupling expansion in $1/\theta$
corresponds to the perturbative incorporation of string corrections
measured by the string tension $\alpha'$.

The ``dual'' correspondence between fields on AdS and conformal fields 
on Minkowski space-time was made concrete by a proposal in \cite{GKP,W}.
As a scalar model, the conformal field dual to the free Klein-Gordon 
field on AdS has been considered as an exercise in \cite{W}. It is a
\gff\ \cite{G1,J,L}. It follows that the perturbative treatment of the
AdS-CFT correspondence amounts to an expansion around a \gff. 

For this reason, we believe it worth while to reconsider the properties 
of \gff s. Generalized free fields have been introduced by Greenberg
\cite{G1} as a new class models for local quantum fields, and have
been further studied by Licht \cite{L} as candidates for more general
asymptotic fields as required by the LSZ asymptotic condition. They
can be characterized in several equivalent ways: the commutator is a
numerical distribution; the truncated (connected) $n$-point functions
vanish for $n\neq 2$; the correlation functions factorize into 2-point
functions; the generating functional for the correlation functions is
a Gaussian. But in distinction from a canonical free field, a \gff\ is
not the solution to an equation of motion, and its 2-point function is
not supported on a mass shell. It rather has the form of a superposition 
\begin{equation} 
\langle\Omega,\;\varphi(x)\varphi(x')\Omega\rangle = 
\int_{\RR_+} d\rho(m^2) \; W_m(x-x') 
\end{equation}
where $W_m$ are the 2-point functions of the Klein-Gordon fields of mass 
$m$ in $d$-dimensional Minkowski space-time. The (positive and polynomially 
bounded) weight $d\rho(m^2)$ occurring in the superposition is known as the 
\klw\ \cite{KL}.

In fact, this is the most general form of the 2-point function of any 
scalar quantum field \cite{KL}, and a general theorem \cite{G2} states that
a field whose 2-point function is supported within a finite interval
of masses, is necessarily a (generalized) free field.   

There are several ways by which \gff s arise. E.g., from every Wightman 
field $\phi$, one can obtain Wightman fields $\phi^{(N)}$, $N\in \NN$, 
as the normalized sum of replicas of $\phi$ on $\HH^{\otimes N}$. This
suppresses the truncated $n$-point functions with a factor $N^{(2-n)/2}$. 
In the ``central'' limit $N\to\infty$, one obtains a \gff\ which 
has the same 2-point function as the original (interacting) field. 
Similarly, in the large $N$ limit of $O(N)$ or $U(N)$ symmetric
theories, all truncated functions of gauge invariant (composite)
fields exhibit a leading factor of $N$, so that if the 2-point
function is normalized, the higher truncated functions are also
suppressed by inverse powers of $N$, and the limit is again a \gff\
\cite{tH2}.%
\footnote{On the other hand, the limit $W$ of $1/N$ times the
  connected functional for finite $N$, $W_N=\log Z_N$, is finite and
  non-Gaussian. But $W$ does not define a quantum field theory of its
  own because it violates positivity (unitarity). The significance of
  this quantity is that $N\cdot W$ gives the asymptotic behavior of the large
  $N$ expansion of $W_N$.}    

Another obvious way to obtain a \gff\ is to restrict a free
Klein-Gordon field of mass $M$ in $1+4$ dimensions to the
$1+3$-dimensional hypersurface $x^4=0$. The resulting \klw\ is 
$d\rho(m^2) = dm^2/\sqrt{m^2-M^2}$, supported at $m^2\geq M^2$.

Finally, the AdS-CFT correspondence associates with the free Klein-Gordon 
field on $d+1$-dimensional Anti-deSitter space-time yields a Gaussian
conformal field in $d$-dimensional Minkowski space-time whose scaling
dimension $\Delta = \frac d2 + \nu$ depends on the Klein-Gordon mass
$M$ through the parameter $\nu=\frac12\sqrt{d^2+4M^2}$. Its 2-point
function proportional to $(-(x-x')^2)^{-\Delta}$ is a superposition of
{\em all} masses with \klw\ $d\rho(m^2) = dm^2 m^{2\nu}$ (cf.\ Sect.~4).  

As a consequence of the continuous superposition of masses, there is no 
Lagrangean and no canonical \set\ associated with a \gff. The first
purpose of this article (entirely unrelated to AdS) is the construction 
of a non-canonical \set\ (Sect.~3). This \set\ turns out to be more
singular than a Wightman field, but it fulfills the requirements as a
density for the generators of (global) space-time symmetries. If
smeared with a test function, it has finite expectation values but
infinite fluctuations in almost every state of the Hilbert space
(including the vacuum). Technically speaking, it is a quadratic form
on the Wightman domain, rather than an unbounded operator. Its
commutator with field operators, however, is well defined as an operator.  

Our construction of the \set\ relies on the fact that on the vacuum
Hilbert space of a \gff\ $\varphi$ there exists a large class of
mutually local%
\footnote{commuting with each other at spacelike distance} 
Wightman fields (including $\varphi$ itself, hence relatively local
w.r.t.\ $\varphi$) which is much larger \cite{Br,G1} than the class of
Wick polynomials of $\varphi$. They form what is known as the Borchers 
class \cite{Bo1,J} of $\varphi$. These fields can be expressed in
terms of $\varphi$ (and its Wick polynomials) by the use of highly  
non-local (pseudo) differential operators or convolutions, while still
satisfying local commutativity with $\varphi$ (i.e., the commutator
vanishes at spacelike separation). This fact illustrates the
distinction between the algebraic notion of local commutativity 
(Einstein locality) underlying the concept of the Borchers class, and
a notion of ``expressibility in terms of local operations''. E.g., in
perturbation theory the Lagrangean interaction density should be local
in the former sense in order to ensure locality of the interacting field.

The second prominent issue of this article is the ``holographic''
identification of a quantum field on AdS (the free Klein-Gordon field
$\phi$) in the Borchers class of the boundary \gff\ $\varphi$ (Sect.~4). 
More precisely, AdS is regarded conveniently as a warped product of
Minkowski space-time with $\RR_+$ (whose coordinate we call $z>0$). 
Then for every fixed value $z$, $\phi(z,\cdot)$ is a Wightman field
$\varphi_z(\cdot)$ on Minkowski space-time in the Borchers class of 
$\varphi$, obtained from $\varphi$ by a non-local (pseudo)
differential operator involving a $z$-dependent Bessel function.   

Local commutativity of the Klein-Gordon field $\phi$ on AdS implies
that the fields $\varphi_z$ and $\varphi_{z'}$ in Minkowski space-time
satisfy a certain ``bonus locality'' (local commutativity at finite
timelike distance). We shall explicitly derive this property as a
consequence of the specific non-local operations relating the AdS 
field to the boundary field, invoking a nontrivial identity for
Bessel functions.   

The canonical \set\ of the Klein-Gordon field on AdS is identified as
a $z$-dependent generalized Wick product of the boundary field. 
Integrating this field over $z$, yields the singular \set\ of the
\gff\ mentioned above. This complies with the fact that the canonical
AdS \set\ is a density in AdS, while the \set\ for the \gff\ on the
boundary is a density in Minkowski space. 

With these findings, we want to point out that \gff s are rather
well-behaved Wightman fields, which moreover are ``closer'' to 
interacting quantum fields than free Klein-Gordon fields. It might be
advantageous to perform a perturbation around a \gff, which has
already the correct 2-point function of the interacting field, while
the perturbation only affects the higher truncated correlations. 

As was noticed implicitly, e.g., in \cite{BDHM}, and systematically
analysed in \cite{DR}, the perturbative approach to the AdS-CFT
correspondence may be understood as a perturbation around a canonical
field on AdS with subsequent restriction to the boundary. The
interaction part of the action is an integral over AdS of some Wick
polynomial in the AdS field. Expressing the latter in terms of the
limiting \gff\ on the boundary, and performing the (regularized)
$z$-integral, one obtains (at least formally) a Lagrangean density on
the boundary which is a generalized Wick polynomial of the boundary \gff.  

In view of this observation, the present work is also considered as a
starting point for a perturbation theory of the \gff\ with generalized
Wick polynomials as interactions, which includes the perturbative AdS-CFT 
correspondence as a special case.

In the last, somewhat tentative section, we point out the relation
between the existence of relatively local fields beyond the Wick
polynomials, and the violation of the time-slice property (primitive
causality \cite{HS}) and ``Haag duality'' for \gff s in Minkowski
space-time. These issues are discussed in terms of the von Neumann
algebras of localized observables associated with a quantum field
\cite{HK}. Although logically unrelated to AdS-CFT, they may be nicely
understood in terms of geometric properties of AdS and its boundary,
using the above holographic interpretation. The discussion also exhibits 
a slight but important difference between the present holographic
picture and the ``algebraic'' notion of holography \cite{KHR}.

\section{Generalized free fields}
\setcounter{equation}{0} 
Let $\MM^d=(\RR^d,\eta_{\mu\nu})$ denote $d\geq 2$-dimensional Minkowski 
space-time, and $V_+$ the open forward light-cone (in momentum space). 

We consider a hermitean scalar \gff\ \cite[Chap.\ 2.6]{J} on $\MM^d$
with \klw\ $d\rho(m^2) = dm^2$ on $\RR_+$. It has the form
\begin{equation} 
\varphi(x) = \int_{V_+} d^dk \; [a(k) e^{-ikx} + a^+(k)e^{ikx}] 
\end{equation}
in terms of creation and annihilation operators 
\begin{equation} 
[a(k),a^+(k')] = (2\pi)^{-(d-1)}\delta^d(k-k'),
\qquad [a,a] = 0 = [a^+,a^+]. 
\end{equation}
It is defined on the Fock space $\HH$ over the 1-particle space 
$\HH_1 = L^2(V_+,d^dk)$, identifying 
\begin{equation} 
L^2(V_+,d^dk) \ni f \quad \equiv\quad 
(2\pi)^{\frac {d-1}2} \int_{V_+} d^dk \; f(k) a^+(k)\Omega \in \HH_1. 
\end{equation} 
$\HH_1$, and hence $\HH$, is equipped with the obvious unitary 
positive-energy representation of the \poi\ group with generators%
\footnote{The reader might be worried about the meaning of derivatives
  of $a(k)$. The expressions (2.4), (2.5) as well as (2.17), (2.22)
  below are understood in the distributional sense, i.e., after
  application of an integral $\int d^dk a^+(k)Xa(k)$ to a 1-particle
  vector of the form (2.3) the differential operator $X$ is found
  acting on the smearing function $f \in L^2(V_+,d^dk)$, and likewise
  for $n$-particle vectors. Thus, (2.4), (2.5), (2.17), (2.22) are the
  ``second quantizations'' of the corresponding differential operators
  on $L^2(V_+,d^dk)$. We shall not discuss here the precise domains on
  which these hermitean generators are (essentially) self-adjoint.}
\begin{align} 
P_\mu &= (2\pi)^{d-1} \int_{V_+} d^dk \;  a^+(k) \; k_\mu \; a(k), \\ 
M_{\mu\nu} &= (2\pi)^{d-1} \int_{V_+} d^dk \; 
a^+(k)\; i(k_\nu\frac{\partial}{\partial k^\mu} - 
k_\mu\frac{\partial}{\partial k^\nu}) \; a(k),  
\end{align}
such that
\begin{align} 
i[P_\mu,\varphi(x)] &= \partial_\mu\varphi(x), \\
i[M_{\mu\nu},\varphi(x)] &= (x_\mu\partial_\nu-x_\nu\partial_\mu)
\varphi(x). 
\end{align}
The \gff\ $\varphi$ is a local field because its commutator reads
\begin{eqnarray} 
[\varphi(x),\varphi(x')] = (2\pi)^{-(d-1)} \int_{V_+} d^dk \;
\big(e^{ik(x-x')}-e^{-ik(x-x')}\big) = \qquad \nonumber 
\\ = \int_{\RR_+} dm^2 \; (2\pi)^{-(d-1)} \int_{V_+} d^dk \; 
\delta(k^2-m^2) \; \big(e^{ik(x-x')}-e^{-ik(x-x')}\big) \nonumber \\[-4pt] 
= \int_{\RR_+} dm^2 \Delta_m(x-x') 
\end{eqnarray}
where $\Delta_m$ is the commutator function of the free Klein-Gordon field 
of mass $m$.%
\footnote{The Klein-Gordon fields themselves are not present in the
  theory, though,  because square integrable functions in $\HH_1$
  cannot have sharp mass.} 

\subsection{Relatively local \gff s and generalized Wick products}

Our first observation is that on the same Hilbert space, we can define
\begin{equation} 
\varphi_h(x)= \int_{V_+} d^dk \; h(k^2) \; [a(k) e^{-ikx} + a^+(k)e^{ikx}] 
\end{equation}
with $h$ any smooth polynomially bounded real function on $\RR_+$ 
(called ``weight function''). These are again hermitean scalar fields
on $\HH$, satisfying (2.6) and (2.7). Moreover, all $\varphi_h$ are
local and mutually local fields, because their commutators 
\begin{equation} 
[\varphi_{h_1}(x),\varphi_{h_2}(x')] = \int_{\RR_+} dm^2 
\; h_1(m^2)h_2(m^2) \; \Delta_m(x-x')
\end{equation}
vanish at spacelike distance irrespective of the functions $h_i$.
By inspection of the 2-point functions
\begin{equation} 
\langle\Omega, \varphi_{h_1}(x) \varphi_{h_2}(x') \Omega\rangle = 
\int_{\RR_+} dm^2 \; h_1(m^2)h_2(m^2) \; W_{m}(x-x'), 
\end{equation}
one sees that each $\varphi_h$ is a \gff\ with \klw\ $dm^2 h(m^2)^2$. 
In fact, the weight function $h$ need not be smooth as long as 
$dm^2\,h(m^2)^2$ is a polynomially bounded measure.

If the weight function $h$ is a polynomial, then 
\begin{equation} 
\varphi_h = h(-\square)\varphi 
\end{equation}
is just a derivative of $\varphi$, and 
$\varphi_h(f) = \varphi(h(-\square)f)$ where the support of
$h(-\square)f$ equals (a subset of) the support of $f$. Hence
$h(-\square)$ is a local operation. But if $h$ is not a polynomial,
then $h(-\square)$ may be tentatively defined on $f$ by multiplication
of the Fourier transform $\hat f(k)$ with any function of $k^2$ which
coincides with $h$ on $\RR_+$ (all giving the same field operator
$\varphi(h(-\square)f)$). This is a highly non-local operation which
does not preserve supports. Likewise, one may formally read (2.9) as a
convolution in $x$-space \cite{G1}    
\begin{equation}
\varphi_h(x) = \int_{\MM^d}d^dx\; \check H(x-y)\varphi(y)
\end{equation}
with the distributional inverse Fourier transform of any function $H(k)$ 
which equals $h(k^2)$ on $V_+$. E.g., if $h$ is analytic, $h(-\square)$ 
and $H(k)$ may be defined as power series; but as the example of
$h(z)=\cos\sqrt z$ exemplifies, the inverse Fourier transform of
$h(k^2)\hat f(k)$ or $H(k)$ may not exist due to the rapid growth of
$h(k^2)$ at negative $k^2$. Therefore, expressions like (2.12) or
(2.13) in the general case should not be taken literally. These are
suggestive ways of rewriting the definition (2.9), indicating a
non-local operation on $\varphi$ which yet gives rise to a local field.    

The above construction of fields satisfying local commutativity with 
$\varphi$ and among themselves can be extended to Wick products
\cite{Br,G1}. The expressions 
\begin{eqnarray} 
(:\varphi^2:)_h(x)= \int_{V_+} d^dk_1 \int_{V_+} d^dk_2 \;
h(k_1^2,k_2^2) \qquad\qquad \qquad\qquad\qquad \nonumber \\[-3pt] 
:[a(k_1) e^{-ik_1x} + h.c.] [a(k_2) e^{-ik_2x} + h.c.]: 
\end{eqnarray} 
define Wightman fields, relatively local with respect to $\varphi$ and
$\varphi_{h'}$ and mutually local among each other, for every (smooth)
polynomially bounded real symmetric function $h$ on $\RR_+ \times \RR_+$. 
Formally, they may be represented as point-split limits of the form 
\begin{equation} 
(:\varphi^2:)_h(x)= \lim_{x'\to x} h(-\square,-\square')
\big(\varphi(x)\varphi(x') - 
\langle\Omega,\;\varphi(x)\varphi(x')\Omega\rangle\big).
\end{equation}

The smoothness of the functions $h$ may be considerably relaxed. While 
we refer to \cite{Br} for details, we point out that for $(:\varphi^2:)_h$ 
to be a Wightman field, $h^2$ ought to be at least a measurable function on 
$\RR_+ \times \RR_+$: otherwise $(:\varphi^2:)_h(f)$ fails to be an 
operator with the vacuum vector in its domain. This can be seen 
easily from the 2-point function 
\begin{eqnarray} 
\langle\Omega, (:\varphi^2:)_h(x) (:\varphi^2:)_h(x') \Omega\rangle = 
\quad\qquad\qquad\qquad\qquad\qquad \nonumber \\ = 2 \int_{\RR_+} 
dm_1^2 dm_2^2 \; h(m_1^2,m_2^2)^2 \; W_{m_1}(x-x') W_{m_2}(x-x'). 
\end{eqnarray}

It is clear how this construction generalizes to higher Wick polynomials, 
and also to multi-local fields such as $(:\varphi(x_1)\varphi(x_2):)_h$. 
All these fields satisfy local commutativity among each other with
respect to their arguments inspite of the non-local operations
involved. It is crucial that the weight functions $h$ depend only on
the squares of the four-momenta, since general functions of the
components $k^\mu$ would spoil local commutativity. It is also clear
that the construction can be as well applied to Wick polynomials of
derivatives $\partial_\mu\dots\partial_\nu\varphi$ of the \gff. 

Generalized Wick polynomials belong to the Borchers class of the \gff\
consisting of the relatively local Wightman fields defined on the same
Hilbert space. We believe that they exhaust the Borchers class \cite{E}.
They are natural candidates for perturbative interactions, e.g., in
causal perturbation theory \cite{EG,BF2,DF2}.  

\subsection{Conformal symmetry}

The 1-particle Hilbert space $\HH_1$, and hence the Fock space $\HH$, carry 
also a natural representation of the group of dilations with generator
\begin{equation} 
D = (2\pi)^{d-1}  \int_{V_+} d^dk \; a^+(k) \;{\textstyle \frac i2}
\big ((k\cdot\partial_k) + (\partial_k \cdot k) \big) \; a(k).
\end{equation}
Under this representation, the \gff s $\varphi_h$ transform according to
\begin{equation} 
U(\lambda)\varphi_h(x)U(\lambda)^* = \varphi_{h_\lambda}(\lambda x) 
\end{equation}
where $h_\lambda(m^2) = \lambda^{\frac d2}h(\lambda^2m^2)$. In particular, 
the \gff s with homogeneous weight functions $m^\nu$, 
\begin{equation} 
\varphi^{(\Delta)} = (-\square)^{\nu/2} \varphi 
\qquad \hbox{with} \quad \nu \equiv \Delta - {\textstyle \frac d2}
\end{equation}
transform like scale-invariant fields of scaling dimensions 
$\Delta = \frac d2+\nu$: 
\begin{equation} 
U(\lambda)\varphi^{(\Delta)}(x)U(\lambda)^* = 
\lambda^\Delta\varphi^{(\Delta)}(\lambda x), 
\end{equation}
or in infinitesimal form 
\begin{equation} 
i[D,\varphi^{(\Delta)}(x)] = 
(x^\mu\partial_\mu + \Delta)\varphi^{(\Delta)}(x). 
\end{equation}
$U(\lambda)$ scales the momenta (2.4) and commutes with the Lorentz
transformations (2.5). Hence it extends the representation of the \poi\ 
group to a representation of the \poi-dilation group, also denoted by $U$.

It is well known, that the scale-invariant 2-point functions of these fields 
are in fact conformally covariant, extending the fields to a suitable
covering of Minkowski space-time $\CM^d$ \cite{LM}. This means in
particular that the unitary representation $U$ of the \poi-dilation group 
extends to a unitary representation $U^{(\Delta)}$ of the conformal
covering group. Under $U^{(\Delta)}$, the field $\varphi^{(\Delta)}$
transforms as a conformally covariant scalar field of scaling dimension 
$\Delta$. Unlike those of the \poi-dilation group, the generators 
of the special conformal transformations depend on the parameter
$\Delta = \frac d2 +\nu$ and are explicitly given on $\HH_1$ by  
\begin{eqnarray} 
K^{(\Delta)}_\mu =  (2\pi)^{d-1} \int_{V_+} d^dk 
\qquad\qquad\qquad\qquad\qquad\qquad\qquad\qquad \nonumber \\
a^+(k) \; \bigg(\frac\partial{\partial k^\alpha}
k_\mu\frac\partial{\partial k_\alpha} - (k\cdot\partial_k)
\frac\partial{\partial k^\mu} - \frac\partial{\partial k^\mu}  
(\partial_k \cdot k) + \nu^2 \frac{k_\mu}{k^2} \bigg) \; a(k), 
\end{eqnarray}
such that
\begin{equation} 
i[K^{(\Delta)}_\mu,\varphi^{(\Delta)}(x)] = \big( 2x_\mu(x\cdot\partial) 
- x^2\partial_\mu + 2\Delta x_\mu \big) \; \varphi^{(\Delta)}(x). 
\end{equation}

We emphasize that although the fields $\varphi^{(\Delta)}$ are for all
values of $\Delta$ defined on the same Hilbert space (the common Fock
space $\HH$ for all \gff s $\varphi_h$), they are conformally
covariant with respect to different representations $U^{(\Delta)}$ of
the conformal group on the same Hilbert space. These representations
coincide only on the \poi-dilation subgroup. $U^{(\Delta)}$ does not
implement a geometrical point transformation of $\varphi^{(\Delta')}$,
$\Delta' \neq \Delta$, nor of $\varphi_h$ in general.

\section{The stress-energy tensor}
\setcounter{equation}{0} 
The purpose of this section is to find a \set\ $\Theta_{\mu\nu}(x)$
for the \gff\ (2.1) which has the properties of a local and covariant 
conserved tensor density for the generators of the \poi\ group. 
It should thus satisfy 
\begin{align} 
\partial^\mu\Theta_{\mu\nu} &= 0, 
\\ \int d^{d-1}\vec x \; \Theta_{0\nu} &= P_\nu, \\[-4pt] 
\int d^{d-1}\vec x \; (x_\mu\Theta_{0\nu} - x_\nu\Theta_{0\mu}) 
&= M_{\mu\nu}, 
\\ \Theta_{\mu\nu}(x) &= \Theta_{\nu\mu}(x), \\ {} 
[\Theta_{\mu\nu}(x),\varphi_h(x')] &= 0 \qquad ((x-x')^2<0). 
\end{align}

With an ansatz of the form (2.14), including derivatives of $\varphi$,
we find the solution
\begin{equation} 
\Theta_{\mu\nu} = \big(:\partial_\mu\varphi\partial_\nu\varphi - 
{\textstyle \frac 12}
\eta_{\mu\nu} \; (\partial_\alpha\varphi\partial^\alpha\varphi + 
\varphi\square\varphi):\big)_\delta. 
\end{equation}
The generalized Wick product $(:\dots:)_\delta$ is understood as
$(:\dots:)_h$ with the choice of the weight ``function'' 
\begin{equation} 
h(m_1^2,m_2^2) = \delta(m_1^2,m_2^2) \equiv \delta(m_1^2 - m_2^2)
\theta(m_1^2). 
\end{equation}

This singular choice cannot be avoided due to the requirement (3.2),
since the spatial integral over a generalized Wick square such as 
(2.14) can only enforce equality of the spatial components of the
momenta of the creation and annihilation operators, $\vec k_1=\vec k_2$, 
while the representation (2.4) of the total momentum operator requires
$k_1^\mu = k_2^\mu$. But with this singular weight function $h$, the
2-point function (2.16) involving $h^2$ becomes highly divergent,
hence the \set\ has infinite fluctuations in the vacuum state. Smearing 
$\Theta_{\mu\nu}$ with a test function does not give an operator whose
domain contains the vacuum vector. Thus the \set\ is not a Wightman field.  

But $\Theta_{\mu\nu}(f)$ is a quadratic form on the Wightman domain of
$\varphi_h$, i.e., its matrix elements with vectors from that domain are 
finite (more precisely: are continuous functionals of the test function $f$). 
To prove this, only the finiteness of 
\begin{eqnarray} 
\langle\Omega,\; 
\Theta_{\mu\nu}(f)\varphi_{h_1}(f_1)\varphi_{h_2}(f_2)\Omega\rangle, \nonumber
\\ \langle\Omega,\; 
\varphi_{h_1}(f_1)\Theta_{\mu\nu}(f)\varphi_{h_2}(f_2)\Omega\rangle, 
\\ \langle\Omega,\;
\varphi_{h_1}(f_1)\varphi_{h_2}(f_2)\Theta_{\mu\nu}(f)\Omega\rangle \nonumber
\end{eqnarray}
needs to be checked since every matrix element of $\Theta_{\mu\nu}(f)$ on
this domain is a sum of terms of either of these forms with finite 
coefficients. Explicit evaluation of the above matrix elements, which are
all of the form
\begin{eqnarray} 
\int_{V_+} d^dk_1 \; \hat f_1(\pm k_1)h_1(k_1^2)
\int_{V_+} d^dk_2 \; \hat f_2(\pm k_2)h_2(k_2^2) \; \delta(k_1^2-k_2^2) 
\quad \nonumber \\[-4pt]  \times P(k_1,k_2)\hat f({}\pm k_1 \pm k_2) 
\end{eqnarray}
with $P$ some polynomial, exhibits their finiteness and continuity with 
respect to $f$, for arbitrary test functions $f_i$ and arbitrary weight 
functions $h_i$. $\Theta_{\mu\nu}(f)$ being a quadratic form on the
Wightman domain of $\varphi_h$, its commutator with $\varphi_h(g)$ is
a priori well defined as a quadratic form. It turns out to be in fact
an operator on the Wightman domain. It vanishes if the supports of $f$
and $g$ are spacelike separated. 

The formula (3.6) for $\Theta_{\mu\nu}$ is uniquely determined by the 
requirements (3.1--5), up to addition of a multiple of
$(\partial_\mu\partial_\nu-\eta_{\mu\nu}\square) (:\varphi^2:)_\delta$.
We note that $\Theta_{\mu\nu}$ is not traceless, nor can it be made traceless
by such an addition. Thus it does not provide a density for the generator 
of the dilations (2.17), nor for the conformal transformations (2.22).

\section{Application to AdS-CFT}
\setcounter{equation}{0} 

The conformal group $SO(2,d)$ of $d$-dimensional Minkowski space-time
coincides with the group of isometries of $d+1$-dimensional anti-deSitter
space-time $\ADS_{d+1}$. In fact, $\ADS_{d+1}$ has a conformal
boundary which is a twofold covering%
\footnote{We denote by $\ADS_{d+1}$ the quadric $\xi\cdot\xi=1$ in
  $\RR^{2,d}$ (signature $(+,+,-,\dots,-)$), and by $\PADS_{d+1}$ its
  quotient by the antipodal identification $\xi \leftrightarrow-\xi$. 
  The conformal boundary of this quotient is the Dirac
  compactification $\CM^d$ of Minkowski space-time whose points are
  the lightlike rays $\zeta=\RR\cdot n$, $n \cdot n =0$, in $\RR^{2,d}$. 
  The field theories discussed below in general are defined on
  covering spaces of the respective manifolds.}   
of the Dirac compactification $\CM^d$ of Minkowski space-time, such that 
the AdS group restricted to the boundary acts like the conformal group 
on $\CM^d$. One-parameter subgroups with future-directed timelike tangent 
vectors in AdS (``time evolutions'') have future-directed timelike tangent 
vectors in $\CM^d$. Hence, the respective (AdS and conformal) notions
of ``positive energy'' for the unitary representations of $SO(2,d)$ coincide.

It was shown in \cite{F} that the scalar Klein-Gordon field on $\ADS_{d+1}$ 
can be canonically quantized on a 1-particle space which carries a 
positive-energy representation of the AdS group.

Our aim is to find the explicit relation between these scalar Klein-Gordon 
fields $\phi$ on $\ADS_{d+1}$ (parametrized by a parameter $\nu>-1$ such 
that the Klein-Gordon mass equals $M^2 = \nu^2 - \frac{d^2}4$) and the 
\gff\ on Minkowski space-time $\MM^d$ (characterized by its scaling dimension 
$\Delta = \frac d2+\nu$). Both fields are defined on the same Fock space $\HH$ 
over the 1-particle space $\HH_1$ (2.3), carrying the same unitary
positive-energy representation $U^{(\Delta)}$ of $SO(2,d)$ under which both 
fields transform covariantly in the respective (AdS or conformal) sense. 

We shall work in the convenient chart of AdS given by \poi\
coordinates $x^M \equiv (z\in\RR_+,x^\mu \in \MM^d)$, in which the
metric takes the form  
\begin{equation} 
ds^2 = g_{MN}\; dx^Mdx^N = z^{-2}\cdot(\eta_{\mu\nu}\;dx^\mu dx^\nu -dz^2),
\end{equation}
i.e., it is a ``warped product'' of $d$-dimensional Minkowski space-time 
$\MM^d$ by $\RR_+$, or in the terminology of \cite{BBMS}, AdS has a
foliation by $\MM^d$. The chart is given as follows. We fix a pair
$e_\pm$ of lightlike vectors in $\RR^{2,d}$, $e_+\cdot e_-=\frac12$,
and a basis $e_\mu$ of the subspace orthogonal to $e_\pm$, with 
$e_\mu\cdot e_\nu=\eta_{\mu\nu}$. Then  
\begin{equation}
\xi=z\inv\cdot(x^\mu e_\mu + e_- + (z^2-x_\mu x^\mu)e_+)
\end{equation}
fulfills $\xi\cdot\xi=1$. This chart covers $\PADS_{d+1}$ except for
the hypersurface $\xi\cdot e_+ = 0$ which formally corresponds to
$z=\infty$.  

The corresponding chart $(x^\mu\in\MM^d)$ of $\CM^d$, parametrizing the
lightlike rays in $\RR^{2,d}$ by  
\begin{equation}
\zeta = \RR\cdot(x^\mu e_\mu + e_- -x_\mu x^\mu e_+),
\end{equation}
is Minkowski space-time $\MM^d\subset\CM^d$. This chart misses out the
hypersurface of compactification points ``at infinity'' of $\MM^d$,
consisting of the lightlike rays orthogonal to $e_+$, namely the rays 
$\RR\cdot(\lambda e_++x^\mu e_\mu)$, $x_\mu x^\mu=0$. In $\CM^d$,
these are the points at lightlike distance from
$\omega\equiv\RR\cdot e_+$ (the compactification point
``at spacelike infinity'' of $\MM^d$). 

$\zeta(x^\mu)$ is the boundary point approached by $\xi(z,x^\mu)$ as
$z\to 0$. Thus the \poi\ chart meets the boundary exactly in
Minkowski space-time. 

From this discussion, we conclude that the chart $(z,x^\mu)$ is mapped
onto itself only by the stabilizer subgroup in $SO(2,d)$ of the ray
$\RR_+\cdot e_+$ in $\RR^{2,d}$. This subgroup has the form 
\begin{equation}
(SO(1,d-1) \times \RR_+)\ltimes\RR^d.
\end{equation}
Here, $SO(1,d-1) \ltimes\RR^d$ is the stabilizer group of the vector $e_+$. 
$SO(1,d-1)$ preserves also $e_-$ and transforms the basis $e_\mu$ like
the Lorentz group, while $\RR^d$ takes $e_-\mapsto e_-+a^\mu e_\mu
-a_\mu a^\mu e_+$, $e_\mu \mapsto e_\mu-2a_\mu e_+$. Hence the stabilizer 
subgroup of $e_+$ preserves the coordinate $z= (2\xi\cdot e_+)\inv$
and acts on $x^\mu$ like the \poi\ group. The remaining factor $\RR_+$
in (4.4) scales $e_\pm\mapsto\lambda^{\pm 1}e_\pm$ and preserves $e_\mu$, 
hence it takes $(z,x^\mu)$ to $(\lambda z,\lambda x^\mu)$ and acts on the
boundary $z=0$ like the dilations. We shall refer to these subgroups of
$SO(2,d)$ as \poi\ and dilation subgroups also in the AdS context. 
Thus, the \poi\ chart of AdS is preserved by the \poi-dilation group
of Minkowski space-time. The remaining elements of $SO(2,d)$ induce
rational transformations of the coordinates $(z,x^\mu)$, such as 
$(z,x^\mu) \mapsto (z,x^\mu-b^\mu(x^2-z^2))/(1-2(b\cdot x)+b^2(x^2-z^2))$,
$b \in \MM^d$, restricting to the special conformal transformations of
the boundary $z=0$.

\subsection{The Klein-Gordon field on AdS}

We fix any value $\nu > -1$ and set $\Delta = \frac d2 + \nu$
and $M^2 = \Delta(\Delta-d) = \nu^2 - \frac{d^2}4$. 

The Klein-Gordon field on AdS 
\begin{equation} 
(\square_g + M^2)\phi = (-z^{1+d}\partial_zz^{1-d}\partial_z + z^2
\square_\eta + M^2) \phi = 0 
\end{equation} 
has been quantized with an AdS-invariant vacuum state, e.g., in
\cite{AIS,F}. Its 2-point function can be displayed in the form \cite{BBMS} 
\begin{eqnarray} 
\langle\Omega,\;\phi(z,x)\phi(z',x')\Omega\rangle = 
{\textstyle \frac12} (zz')^{\frac d2} \int_{\RR_+} dm^2 \; 
J_\nu(zm)J_\nu(z'm) \; W_m(x-x') \nonumber  \\ = 
(2\pi)^{-(d-1)} {\textstyle \frac12}(zz')^{\frac d2} \int_{V_+} d^dk \; 
J_\nu(z\sqrt{k^2}) J_\nu(z'\sqrt{k^2}) \; e^{-ik(x-x')}. 
\end{eqnarray}
Here, $J_\nu$ is the Bessel function solving Bessel's differential equation
\begin{equation} 
((u\partial_u)^2+u^2) J_\nu(u) = \nu^2 J_\nu(u), 
\end{equation}
and $z^{\frac d2}J_\nu(z\sqrt{k^2})e^{\pm ikx}$ ($k\in V_+$) are the 
plane-wave solutions of the Klein-Gordon equation (4.5). 

We note that, depending on the integrality of the parameter 
$\Delta=\frac d2 + \nu$, the quantum field $\phi$ is in general defined on 
a covering space of $\ADS_{d+1}$ \cite{BBMS,F}. This complicates the
analysis, but the complications precisely match the complications
arising in the corresponding conformal QFT which is defined on a
covering space of $\CM^d$. We shall limit ourselves to the
Klein-Gordon field on the \poi\ chart $(z,x^\mu)$, and correspondingly
to boundary fields in Minkowski space-time.

We also note that for $\vert\nu\vert < 1$, the two possible signs of $\nu$
give rise to inequivalent covariant quantizations \cite{BF1} (in fact,
an interpolating one-parameter family \cite{BBMS}) of the Klein-Gordon
field of the same mass. The commutator functions derived from (4.6)
are the same for both signs of $\nu$ \cite{BBMS}, hence both quantum
field theories have the same {\em local} structure. 

$\phi$ satisfies the canonical equal-time commutation relation between the
field and its canonical momentum $\pi = z^{1-d}\partial_0\phi$
\begin{equation} 
[\phi(z,x),\pi(z',x')]\vert_{x^0=x'{}^0} = i\delta^{d-1}(\vec x-\vec x') 
\delta(z-z'),  
\end{equation}
as can be verified from (4.6), using the fact that $W_m(x-x')$ satisfies 
canonical commutation relations on $\MM^d$, and using Hankel's identity
\begin{equation} 
\int_0^\infty t\,dt \; J_\nu(tu)J_\nu(tu') = u\inv \;
\delta(u-u'), 
\end{equation} 
which expresses the completeness of the plane-wave solutions involved in the 
integral (4.6). 

\subsection{Expression in terms of \gff s}

By comparison of (4.6) with (2.11), we conlude that $\phi(z,x)$ can be
identified with
\begin{equation} 
\phi(z,x) = {\textstyle \frac 1{\sqrt 2}} \; z^{\frac d2} 
\int_{V_+} d^dk \; J_\nu(z\sqrt{k^2}) [a(k) e^{-ikx} + a^+(k)e^{ikx}].
\end{equation}
This is of the form 
\begin{equation} \phi(z,x) = \varphi_{h_z}(x) \end{equation} 
with
\begin{equation} 
h_z(m^2) = {\textstyle \frac 1{\sqrt 2}} \; z^{\frac d2} J_\nu(zm) 
\end{equation} 
i.e., for each value of $z$, $\phi(z,\cdot)$ is one of the \gff s
considered in Sect.\ 2.1.

From the power law behaviour of 
$J_\nu(u) \approx \frac{2^{-\nu}}{\Gamma(\nu+1)} \; u^\nu (1 + O(u^2))$ 
at small arguments, one obtains
\begin{equation} 
\lim_{z\to 0} z^{-\Delta} \phi(z,x) = 
\textstyle{\frac{2^{-\nu-\frac 12}}{\Gamma(\nu+1)}} \; 
\varphi^{(\Delta)}(x), 
\end{equation}
i.e., the \gff\ $\varphi^{(\Delta)}$ is the boundary limit of the
Klein-Gordon field on AdS. This agrees with the results discussed in
\cite{BBMS} and \cite{DR}. 

\subsection{Identification of the representations of $SO(2,d)$}

The 2-point function (4.6) is AdS-invariant, i.e., it depends only on
the AdS-invariant ``chordal distance'' (the distance within
$\RR^{2,d}$) $(-(x-x')^2+(z-z')^2)/2zz'$. 
It follows that the representation of the AdS group defined by
\begin{equation} 
U^{(\kg)}(g)\;\phi(z,x)\Omega := \phi(g(z,x))\Omega \qquad (g \in SO(2,d))
\end{equation}
is unitary on the 1-particle space $\HH_1$, and hence on $\HH$. It 
implements the covariant transformation law 
\begin{equation} 
U^{(\kg)}(g)\phi(z,x)U^{(\kg)}(g)^* = \phi(g(z,x)). 
\end{equation}

We want to show that this representation of $SO(2,d)$ coincides with the 
representation $U^{(\Delta)}$ of the conformal group, constructed on
the same Hilbert space by the extension of the scale-invariant field
$\varphi^{(\Delta)}$ on $\MM^d$ to the conformally invariant field  
on $\CM^d$ (c.f.\ Sect.\ 2.2).

Eq.\ (4.15), restricted to the boundary $z=0$ where $g$ acts like the 
conformal group on $x$, shows that $U^{(\kg)}(g)$ implements the same 
conformal point transformation of the limiting field $\varphi^{(\Delta)}$
as $U^{(\Delta)}(g)$. More specifically, consider the infinitesimal
form of the AdS transformation law (4.15) for the relevant subgroups,
\begin{align} 
i[P^{(\kg)}_\mu,\phi(z,x)] &= \partial_\mu\phi(z,x), \nonumber \\
i[M^{(\kg)}_{\mu\nu},\phi(z,x)] &= (x_\mu\partial_\nu - x_\nu\partial_\mu)
\phi(z,x), \nonumber \\
i[D^{(\kg)},\phi(z,x)] &= (z\partial_z + x^\mu\partial_\mu)\phi(z,x),
\nonumber \\
i[K^{(\kg)}_\mu,\phi(z,x)] &= \big(2x_\mu (z\partial_z + (x\cdot\partial)) + 
(z^2 - x^2)\partial_\mu \big)\phi(z,x).
\end{align}
In the limit $z\to 0$ according to (4.13), the right-hand sides of (4.16)
turn into (2.6), (2.7), (2.21), (2.23), respectively. Thus, the 
infinitesimal generators of the respective subgroups coincide. 
We conclude that the two representations $U^{(\kg)}$ and $U^{(\Delta)}$
of $SO(2,d)$ coincide, cf.\ \cite{Do}.

With the help of the representation $U^{(\kg)}=U^{(\Delta)}$, 
the AdS field $\phi$ and the boundary field $\varphi^{(\Delta)}$
extend to the respective covering spaces of $\PADS_{d+1}$ and $\CM^d$. 
We emphasize once more that this does not apply for the other boundary
fields $\varphi_h$. 

\subsection{``Holographic'' interpretation}

Combining (4.11) and (2.19), we find
\begin{equation} 
\phi(z,x) = \textstyle{\frac 1{\sqrt 2}}\; z^\Delta \; 
j_\nu(-z^2\square) \; \varphi^{(\Delta)}(x), 
\end{equation}
where 
\begin{equation} 
j_\nu(u^2) = u^{-\nu}J_\nu(u) 
\end{equation}
is a (polynomially bounded) convergent power series in $u^2$. This 
is an explicit expression for the Klein-Gordon field on AdS in terms of its 
limiting \gff\ on the boundary. This ``holographic'' relation%
\footnote{Our use of the term ``holography'' does not quite match
  the one originally suggested by 't Hooft \cite{tH1}, namely the
  {\em reduction in the bulk} of degrees of freedom of a QFT, ascribed
  to gravitational effects in the presence of a horizon. We rather
  allude to the {\em enhancement on the boundary} of degrees of freedom
  necessary and sufficient to ``encode'' a non-gravitational QFT in
  the bulk.}  
is possible with the help of non-local (pseudo) differential operators of 
the kind discussed in Sect.\ 2. 

Obviously, the fact that $\varphi_h = h(-\square)\varphi$ are mutually 
local fields for arbitrary functions $h$, ensures that $\phi(z,x)$ and
$\phi(z',x')$ written in the form (4.11) or (4.17) commute if $(x-x')^2< 0$. 
But this is less than locality on AdS, which requires that the commutator
must vanish even if $(x-x')^2 < (z-z')^2$. Of course we {\em know}
that $\phi$ is a local field on AdS, so the stronger local
commutativity of \gff s $\varphi_{h_z}(x)$ and $\varphi_{h_{z'}}(x')$
at finite timelike Minkowski distance must be true. 

One can understand the origin of this ``bonus locality'' for the
\gff s involved. Evaluating the commutator function according to (2.10), 
gives an integral over three Bessel functions (because $\Delta_m$ at 
timelike distance $(x-x')^2=\tau^2$ is also given by a Bessel function
$(m/\tau)^{\frac{d-2}2} J_{\frac{2-d}2}(m\tau)$) of the form
\begin{equation} 
I(a,b,c) = \int_0^\infty u^{1-\mu}\,du \; J_\mu(au) J_\nu(bu)
J_\nu(cu) 
\end{equation}
with $a^2 = (x-x')^2$, $b=z$, $c=z'$ and $\mu = \frac{2-d}2$. This integral 
can be found, e.g., in \cite[Sect.\ 13$\cdot$46(1)]{Wa}, where it is shown 
to vanish if $a^2 < (b-c)^2$. Thus we precisely find local commutativity 
on AdS. It is the specific form of the Bessel functions in (4.17), 
(4.18) which is able to ensure locality in a higher-dimensional space-time. 

This remark should make it clear that defining 
$\phi(z,x) = \varphi_{h_z}(x)$ with any suitable family of functions 
$h_z(m^2)$, depending on a parameter $z$ and solving a suitable 
differential equation with respect to $z$, may well produce a quantum 
field on a higher dimensional space-time solving some equation of
motion, but this field will in general not satisfy local commutativity.%
\footnote{Such constructions were proposed in \cite{Mi}.}

\subsection{The stress-energy tensor}

The Klein-Gordon field on AdS has a canonical covariantly conserved \set\ 
given by 
\begin{equation} 
\Theta^\kg_{MN}(z,x) = {}:D_M\phi D_N\phi - {\textstyle \frac 12}g_{MN}
(g^{AB}D_A\phi D_B\phi - M^2\phi^2):. 
\end{equation}
Because of the special form of the AdS metric, the {\em covariant} tensor
continuity equation gives rise to {\em ordinary} continuity equations
for the Minkowski components $\Theta^\kg_{M\nu}$, $\nu\neq z$,
\begin{equation} 
g^{MN}\partial_{N}(z^{1-d}\Theta^\kg_{M\nu})=0. 
\end{equation}
Thus $z^{1-d}\Theta^\kg_{0\nu}$ are densities of conserved quantitites,
which are the generators of the \poi\ subgroup of $SO(2,d)$,
\begin{align} 
P_\mu &= \int_0^\infty dz z^{1-d} \int d^{d-1}\vec x \; 
\Theta^\kg_{0\mu}(z,0,\vec x) , \\ 
M_{\mu\nu} &= \int_0^\infty dz z^{1-d}\int d^{d-1}\vec x \; 
(x_\mu\Theta^\kg_{0\nu}(z,0,\vec x) - x_\nu\Theta^\kg_{0\mu}(z,0,\vec x)).
\end{align} 

In these integrals, we may express $\phi$ in terms of the \gff\ 
$\varphi$ by means of (4.11), thus introducing two $z$-dependent (Bessel)
weight functions, and perform the $z$-integration. Because of the term 
$g^{MN}\partial_M\phi\partial_N\phi$ involving $z$-derivatives, a partial
integration becomes necessary after which Bessel's differential
equation can be used to eliminate all derivatives of the Bessel 
functions. 

After performing these steps on the Minkowski components of 
$\Theta^\kg_{\mu\nu}$, one ends up with a generalized Wick product of
(derivatives of) $\varphi$, whose weight function is the result of
the $z$-integration over the Bessel functions:
\begin{equation} 
\int_0^\infty dz\, z^{1-d} \;\Theta^\kg_{\mu\nu}(z,x) = 
\big(:\partial_\mu\varphi\partial_\nu\varphi - {\textstyle \frac 12}
\eta_{\mu\nu} \; (\partial_\alpha\varphi\partial^\alpha\varphi + 
\varphi\square\varphi):\big)_h
\end{equation}
where
\begin{equation} 
h(m_1^2,m_2^2) = {\textstyle \frac12} \int_0^\infty 
z\,dz \; J_\nu(zm)J_\nu(zm'). 
\end{equation}
Once more using Hankel's identity (4.9) which in this case plays the role 
of an orthonormality relation, the integral can be performed, 
giving $h(m_1^2,m_2^2) = \delta(m_1^2-m_2^2)$. 

We have thus exactly reproduced the singular \set\ found in Sect.\ 2.2,
\begin{equation} 
\Theta_{\mu\nu}(x) = \int_0^\infty dz \,z^{1-d} \; 
\Theta^\kg_{\mu\nu}(z,x). 
\end{equation}
But the origin of its singular weight function appears in an entirely new 
light: it is the result of the ``holographic'' projection of AdS onto
its boundary. Some cutoff in the $z$-integral would smoothen the
resulting weight function (4.25). The smoothened \set\ would still act
as a density for generators which generate the correct transformation
laws on those AdS fields which are causally disconnected from the AdS
region where the cutoff is effective. In terms of the corresponding
\gff s on the boundary, these are \gff s within a restricted region
{\em and} with a restricted set of weight functions $h_z$.

\section{Local algebras}
\setcounter{equation}{0} 
At first sight, our ``holographic'' result of Sect.\ 4.4 does not
quite agree with the algebraic analysis of AdS-CFT in \cite{KHR}.
Let us sketch the situation.

According to (4.11), the AdS fields at the point $(z,x)$ are expressed in 
terms of Minkowski space-time fields at the point $x$. Hence boundary
observables (smeared fields) localized in a region $K \subset \MM^d$
encode all AdS observables localized in the region 
$V(K) = \pi^{-1}(K)\subset \ADS_{d+1}$ where $\pi$ denotes the projection 
$(z,x) \to x$. We may call this feature ``projective holography''. 

On the other hand, the analysis called ``algebraic holography'' 
in \cite{KHR} is based on the identification 
\begin{equation} 
A_\cft(K) = A_\ads(W) \qquad (W=W(K)) 
\end{equation}
of the local algebras $A_\cft(K)$ of boundary observables localized in 
``double-cones'' $K\subset \widetilde\CM{}^d$ (the conformal transforms of 
$K_0 = \{(x^\mu): \vert \vec x \vert < 1 - \vert x^0\vert\}$ in the covering 
space of the Dirac compactification of Minkowski space-time), with 
local algebras $A_\kg(W)$ of AdS observables localized in the ``wedge'' 
regions $W=W(K) \subset \widetilde\ADS_{d+1}$ (the AdS transforms of 
$W_0 = \{(z,x^\mu): \sqrt{z^2+\vec x^2} < 1 - \vert x^0\vert\}$ in the 
covering space of AdS). The wedge $W(K)$ is the causal completion of 
the boundary region $K$. The map $W=W(K)$ is a bijection which preserves 
inclusions, takes causal complements in $\widetilde\CM{}^d$ into causal 
complements in $\widetilde\ADS_{d+1}$, and is compatible with the respective 
actions of the covering group of $SO(2,d)$. 

For $K\subset\MM^d$, the wedge region $W(K)$ extends only to finite 
``depth'' $z$ into AdS and is strictly smaller than $V(K)$, 
which extends to $z=\infty$ and contains points causally disconnected 
from $W(K)$. Hence $A_\ads(V(K))$ is strictly larger than $A_\ads(W(K))$,
and the two notions of ``holography'' cannot be equivalent.

We shall show how this apparent conflict is resolved, although the 
discussion should by no means be considered as rigorous. We shall 
deliberately ignore most of the technical subtleties involved in the
passage between the Wightman axiomatic formulation of QFT (in terms of 
fields, which are unbounded-operator valued distributions) and the 
Haag-Kastler \cite{HK} algebraic formulation (in terms of localized 
observables, which are bounded operators). But we are confident that 
our argument captures correctly the essential features concerning
the ``size'' of von Neumann algebras of local observables associated
with \gff s and with free fields on AdS. 

The general idea for the passage from fields to local algebras
is to define for any open space-time region $O$ the von Neumann algebra
\begin{equation} 
A(O) := \{\phi(f): {\rm supp}\;f \subset O\}'' 
\end{equation}
where $X'$ stands for the algebra of bounded operators on the given 
Hilbert space which commute with (the closures of) all elements of $X$. By 
von Neumann's density theorem, the double commutant $A(O)$ is the weak 
closure of the bounded functions of the unbounded smeared field operators 
(such as $\exp i\phi(f)$ if $\phi(f)$ is self-adjoint). Obviously, the
algebras increase as the regions increase (``isotony''). Although the 
underlying fields are local, it is less trivial \cite{DF1} that local 
algebras of the form (5.2) associated with spacelike separated regions 
mutually commute (``locality''). A covariant transformation law of the 
fields such as (4.15) involves the corresponding transformation of the 
support of test functions, hence the local algebras (5.2) transform in 
the obvious sense under conjugation with the unitary representatives of the 
group (``covariance''). One may imagine that the fields can be
recovered from the algebras by taking suitably regularized limits of
elements of algebras associated with regions shrinking to a point, see
e.g., \cite{FH}. Together with the group of covariance which involves
the time evolution, the algebraic data determine the quantum field theory. 

\subsection{``Algebraic'' vs.\ ``projective holography''}

Applying the prescription (5.2) to the free Klein-Gordon field of mass
$M$ on AdS, we obtain local algebras 
\begin{equation} 
A_\kg(O), \qquad(O \subset \widetilde\ADS_{d+1}). 
\end{equation} 
Applying the same prescription to the single \gff\ $\varphi^{(\Delta)}$
on the conformal completion of Minkowski space-time, we obtain local algebras 
\begin{equation} 
A_\Delta(K), \qquad (K \subset \widetilde\CM{}^d). 
\end{equation}
Applying it to the the entire family of \gff s $\varphi_h$ with arbitrary 
weight functions $h$ on Minkowski space-time, we obtain local algebras 
\begin{equation} 
A_\tot(K), \qquad (K \subset \MM^d). 
\end{equation}

We have the rather obvious inclusions for $K \subset \MM^d$
\begin{equation} 
A_\Delta(K) \subset A_\kg(W(K)) \subset A_\kg(V(K)) \subset A_\tot(K), 
\end{equation}
of which the first reflects the limit (4.13), the second is isotony, and 
the last reflects (4.11). We shall show that in fact 
\begin{equation} 
A_\Delta(K) = A_\kg(W(K)) 
\end{equation}
are proper subalgebras of
\begin{equation} 
A_\kg(V(K)) = A_\tot(K) = A_\Delta^\dual(K). 
\end{equation}
In this formula, the ``dual completion'' \cite{R} is defined as 
\begin{equation} 
A_\Delta^\dual(K) := A_\Delta(K^c)' 
\end{equation}
where $K^c$ is the causal complement of $K$ within $\MM^d$. Fields which
are relatively local to the given field, are among the generators of the 
dual completion. Roughly speaking, the dual completion is the algebraic 
counterpart of the Borchers class in Wightman quantum field theory. 
By a general theorem \cite{BGL}, conformal invariance ensures
``conformal duality''
\begin{equation} 
A_\Delta(K) = A_\Delta(K')' 
\end{equation}
where $K'$ is the causal complement of $K$ within $\widetilde\CM{}^d$; 
but if $K$ is a double-cone within $\MM^d$, then $K^c$ is strictly
smaller than $K'$, and hence $A_\Delta^\dual(K)$ is expected to be strictly 
larger than $A_\Delta(K)$ (violation of Haag duality).

(5.7) is the holographic identification of $A_\kg(W)$ with $A_\Delta(K)$ 
in the sense of (5.1). It is defined globally, and covariant with respect
to $SO(2,d)$. On the other hand, (5.8) shows that the projective notion 
of holography pertains to $A_\tot(K) = A_\Delta^\dual(K)$ instead. 
Projective holography is defined only with respect to the chosen chart
$(z,x^\mu)$, and is covariant only under the \poi-dilation subgroup.

Before we prove (5.7) and (5.8), we note that $A_\tot(K)$ does not
depend on the parameter $\nu$ specifying the scaling dimension of the 
field $\varphi^{(\Delta)}$ and the mass of the corresponding Klein-Gordon 
field $\phi$. Hence, (5.7) can only be true, if $A_\kg(V(K))$ does 
not change if the generating Klein-Gordon field $\phi$ of mass $M^2$ is 
replaced by $\phi'$ of mass $M'{}^2$. Indeed, for different values $\nu$, 
$\nu'$, we have by (4.10) and Hankel's formula (4.9) 
\begin{equation} 
\phi(z,x) = \int_0^\infty z'dz'\; K_{\nu\nu'}(z,z') \phi'(z',x) 
\end{equation}
with the kernel $K_{\nu\nu'}(z,z') = (z/z')^{\frac d2} \int_0^\infty 
m\,dm\; J_\nu(zm)J_{\nu'}(z'm)$. Since this kernel acts on the $z$
coordinate only, it takes test functions supported in $V(K)$ onto test
functions supported in $V(K)$, and hence $A_\kg(V(K)) = A'_\kg(V(K))$.

Let us now turn to (5.7). We invoke a general theorem \cite{A1,Bo2}
of Wightman QFT. Let $O$ be a double-cone and $T$ a timelike hypersurface 
passing through the apices of $O$. Then, in order to generate $A(O)$ 
as in (5.2), rather than smear the field in $O$ it suffices to smear the 
field and all its normal derivatives along $O \cap T$. In the present case,
$O$ is an AdS wedge $W$, $T$ is the boundary, and $O \cap T$ is the
corresponding boundary double-cone $K$. The normal derivatives of $\phi(z,x)$
are of the form $\lim_{z\to 0}\partial_z^N z^{-\Delta}\phi(z,x)$. By (4.17), 
and because $j_\nu$ is a power series in $-z^2\square$, these derivatives 
vanish if $N$ is odd and are proportional to $\square^n\varphi^{(\Delta)}(x)$ 
if $N=2n$. With $(\square\varphi)(f) = \varphi(\square f)$ we conclude
that $A_\kg(W)$ is in fact generated by $\varphi^{(\Delta)}(f)$, 
${\rm supp}\,f \subset K$. This justifies our claim (5.7). 

Now we turn to (5.8). Because the fields generating $A_\tot(K)$ are
relatively local with respect to $\varphi^{(\Delta)}$, we know that
$A_\kg(V(K)) \subset A_\tot(K) \subset A_\Delta(K^c)'\equiv 
A_\Delta^\dual(K)$. The claim is that equality holds.

$A_\Delta(K^c)$ is generated by all $A_\Delta(J)$ where $J$ are
double-cones in $\MM^d$ spacelike separated from $K$. Hence its
commutant is the intersection, running over the same set of $J$, of
algebras $A_\Delta(J)' = A_\Delta(J')$ (by (5.10)) $ = A_\kg(W(J'))$
(by (5.7)). Now, $J$ is spacelike separated from $K$ and belongs to
$\MM^d$ iff $J'$ contains $K$ and the point $\omega=\RR\cdot e_+$ of
$\CM^d$ (spacelike infinity of $\MM^d$, cf.\ Sect.~4), and iff the wedge
$W(J')$ contains $W(K)$ and has $\omega$ as a boundary point
($\omega\in\partial W$). Thus,   
\begin{equation} 
A_\Delta(K^c)' = \bigcap_{\substack{W \supset W(K)\\ \partial W\ni\omega}} 
A_\kg(W).
\end{equation}
We may choose $K=K_0$. The $x^0=0$ Cauchy surface of $V(K_0)$ is
$C_0 = \{(z,0,\vec x): z>0, \vec x^2 <1\}$. Let $W_{\vec e}$ be the wedges 
with Cauchy surface $C_{\vec e} = \{(z,0,\vec x): z>0, (\vec e\cdot
\vec x) > -1\}$, $\vec e \in \RR^{d-1}$, $\vec e^2 = 1$. Each 
$W_{\vec e}$ contains $W(K_0)$ and has $\omega$ as a boundary point,
and every wedge which contains $W(K_0)$ and has $\omega$ as a boundary point,
contains some $W_{\vec e}$. Thus, by isotony, the intersection of
algebras in (5.12) may be taken over $W_{\vec e}$,
\begin{equation} 
A_\Delta(K^c)' = \bigcap_{\vec e} A_\kg(W_{\vec e}).
\end{equation}
Now, because of the Klein-Gordon equation, the field $\phi(z,x)$ is
expressible in terms of its Cauchy data at $x^0=0$, hence 
$A_\kg(W_{\vec e}) = A_\kg(C_{\vec e})$ and $A_\kg(V(K_0)) =A_\kg(C_0)$  
(time-slice property \cite{HS}, see Sect.~5.2 below). The latter
algebras are generated by the canonical $x^0=0$ Klein-Gordon fields
$\phi$ and $\pi$ (cf.\ Sect.~4.1) smeared over the respective regions
of the Cauchy surface. Thus (5.8) is reduced to the claim
\begin{equation} 
\bigcap_{\vec e} A_\kg(C_{\vec e}) = A_\kg(C_0).
\end{equation}
The independence of Cauchy data associated with disjoint regions
entails \cite{A2} 
\begin{equation} 
\bigcap_{\vec e} A_\kg(C_{\vec e}) = 
A_\kg\big(\bigcap_{\vec e} C_{\vec e}\big).
\end{equation}
Thus (5.8) is a consequence of the geometric fact
$\bigcap_{\vec e} C_{\vec e} = C_0$
for $K=K_0$, and by \poi\ and dilation covariance
for all $K\subset \MM^d$.

\subsection{Time-slice property}

Finally, we consider the validity of the ``time-slice property'' (also 
called ``primitive causality'' \cite{HS}, or ``weak additivity'' in
\cite{KHR})  
\begin{equation} 
A(O) = A(C) 
\end{equation}
if $O$ is the causal completion of its Cauchy surface $C$. $A(C)$ is
generated by the fields and their time derivatives smeared over $C$. 
This property holds for canonical free fields \cite{HS}, and we have
just used it in the case of the Klein-Gordon field on AdS. 

The time-slice property does {\em not} hold for generalized free fields
(with $h$ fixed) \cite{HS}. E.g., for the above double-cone $K_0$ the
Cauchy surface is the ball $B_0 = \{(x^0=0,\vec x): \vec x^2<1\}$. 
Smearing the \gff\ $\varphi^{(\Delta)}$ together with its time
derivatives over this surface, tests only the boundary limits of the
corresponding canonical $x^0=0$ Klein-Gordon fields $\phi$ and $\pi$ and
their $z$-derivatives at $\{(z=0,x^0=0,\vec x): \vec x^2<1\}$. Since this 
set does not constitute a Cauchy surface for the wedge $W(K_0)$, the \gff\ 
smeared over the Cauchy surface of $K_0$ generates only a proper subalgebra 
of $A_\Delta(K_0) = A_\kg(W(K_0))$. This violation of the time-slice
property is a necessary feature of algebraic holography \cite{KHR}. 

We want to show that the time-slice property is restored for the dual
completion $A_\Delta^\dual(K) = A_\tot(K)$:
\begin{equation} 
A_\tot(K) = A_\tot(C). 
\end{equation}

Again we may choose $K=K_0$. The algebra $A_\tot(K_0)$ equals $A_\kg(V(K_0))$ 
by (5.8) and hence is generated by the canonical $x^0=0$ Klein-Gordon fields 
$\phi$ and $\pi$ smeared over the Cauchy surface $C_0$ of $V(K_0)$. By 
(4.11), such smearings are smearings of $x^0=0$ \gff s $\varphi_h(0,\vec x)$ 
and their time derivatives over the Cauchy surface $B_0$ of $K_0$. 
Hence the latter generate $A_\tot(K_0)$. This justifies (5.17).
 
It is natural to consider this restoration of the time-slice property 
as being related to the existence of the singular \set\ (3.6) which is
in some weak technical sense associated with the algebras $A_\tot(K)$
but not with $A_\Delta(K)$, and whose integral over a Cauchy surface
generates the causal time evolution. Even though the \set\ is not itself a
Wightman field, it is not too singular to have this desirable dynamical
consequence for the structure of the local algebras associated with
the Wightman fields of the theory.

\section{Conclusion}
\setcounter{equation}{0}

We have studied \gff s from a general perspective and established,
inspite of the non-canonical nature of these fields, the existence of a 
\set\ which serves as a density for the generators of the \poi\ group
as in the canonical framework. We have pointed out that this \set\ is
a mathematical object which is more singular than a Wightman field, 
but can be obtained as a certain limit of generalized Wick products.

We have then studied the free field AdS-CFT--correspondence in the light of
the previous results on \gff s. In particular, we have given an explicit 
``holographic'' formula expressing the Klein-Gordon field on AdS in
terms of \gff s on the boundary. We have identified the above \set\
for \gff s as an integral ``along the $z$-axis of AdS'' over the
canonical Klein-Gordon \set\ $\Theta^\kg_{\mu\nu}(z,\cdot)$ on AdS.

These results should be useful as a starting point for a perturbation theory
of the AdS-CFT--correspondence. If the AdS field is perturbed by a local 
interaction Lagrangean density $L_I(\phi)$ on AdS, it is expected that
the effect on the conformal field (\gff) on the boundary is that of a  
Lagrangean perturbation by the integral ``along the $z$-axis of AdS'' of 
$L_I(\phi(z,\cdot))$. This integral, akin to (4.24), is again of the form 
of a (regular or singular) generalized Wick product of the boundary field.

Further analysis of this issue will be pursued elsewhere.

\vskip5mm

\noindent{\bf Acknowledgments.} This work was supported in part by the 
Deutsche Forschungsgemeinschaft and by the Erwin Schr\"odinger 
International Institute (ESI), Vienna. The authors gratefully acknowledge
interesting discussions with E.\ Br\"uning, C.J.\ Fewster, K.\ Fredenhagen, 
and S.\ Hollands.

\end{document}